\documentclass[journal]{IEEEtran}
\usepackage{cite}

\ifCLASSINFOpdf
  \usepackage[pdftex]{graphicx}
  \graphicspath{{./FIGURES/}}
\else
  \usepackage[dvips]{graphicx}
  \graphicspath{{./FIGURES/}}
\fi
\usepackage[cmex10]{amsmath}
\usepackage{color}

\begin{document}
%
\title{Contact-induced negative differential resistance in short-channel graphene FETs}
%
%
%

\author{Roberto~Grassi,~\IEEEmembership{Member,~IEEE,}
        Tony~Low,
        Antonio~Gnudi,~\IEEEmembership{Member,~IEEE,}\\
        and~Giorgio~Baccarani~\IEEEmembership{Fellow,~IEEE}
\thanks{This work has been supported by the Italian Project PRIN 2008 prot. 2008S2CLJ9 and the EU project GRADE 317839. The authors acknowledge the CINECA Award N. HP10CPFJ69, 2011 for the availability of high performance computing resources and support.}
\thanks{R. Grassi, A. Gnudi, and G. Baccarani are with E. De Castro Advanced Research Center on Electronic Systems (ARCES), University of Bologna, 40136 Bologna, Italy (e-mail: rgrassi@arces.unibo.it).}
\thanks{T. Low is with IBM T.J. Watson Research Center, Yorktown Heights, New York 10598, USA.}
}

\IEEEpubid{\parbox{\textwidth}{\vspace{0.8cm}\copyright~2012 IEEE. Personal use of this material is permitted. Permission from IEEE must be obtained for all other uses, in any current or future media, including reprinting/republishing this material for advertising or promotional purposes, creating new collective works, for resale or redistribution to servers or lists, or reuse of any copyrighted component of this work in other works.}
}


\maketitle

\begin{abstract}
In this work, we clarify the physical mechanism for the phenomenon 
of negative output differential resistance (NDR) in short-channel graphene FETs (GFETs)
through non-equilibrium Green's function (NEGF) simulations and a simpler semianalytical ballistic model that captures the essential physics. This NDR phenomenon is due to a transport mode bottleneck effect induced by the graphene Dirac point in the different device regions, including the contacts. NDR is found to occur only when the gate biasing 
produces an n-p-n or p-n-p polarity configuration along the channel,
for both positive and negative drain-source voltage sweep. 
In addition, we also explore the impact on the NDR effect of 
contact-induced energy broadening in the source and drain regions
and a finite contact resistance.
\end{abstract}


%
\IEEEpeerreviewmaketitle

\section{Introduction}

\IEEEPARstart{G}{raphene} has attracted considerable interest in recent years for applications in analog radio frequency (RF) electronics \cite{MericIEDM08,Liao10,Lin10,Wu11}. The reason lies in the fact that the high carrier mobility and Fermi velocity of graphene could allow device operation up to the THz range of frequencies, while the small on-off current ratio resulting from the zero bandgap, which currently prevents the use of graphene in digital electronics, does not pose a problem in principle for analog applications \cite{Schwierz10}. Indeed, an integrated RF circuit made of graphene devices has already been demonstrated \cite{Lin11}. However, in more general analog circuits, devices with current saturation, i.e. small output conductance $g_d$, are usually required. This is because the intrinsic voltage gain $g_m/g_d$, where $g_m$ is the device transconductance, must be large. Unfortunately, current saturation in graphene devices is not easily obtained due to the lack of a bandgap.

A quasi-saturation of the output characteristics has actually been reported for some experimental long-channel devices \cite{Meric08,Bai11,Han11}. This quasi-saturation is commonly attributed to a charge
``pinch-off'' effect due to the crossing of the quasi-Fermi level
with the channel potential \cite{Meric08,Thiele10,Jimenez11}, but a similar phenomenon has also been predicted in the ballistic limit \cite{Koswatta11,Ganapathi12}.
Recent experiments have shown that not just quasi-saturation ($g_d \rightarrow 0$) but also negative differential resistance ($g_d < 0$) is possible \cite{Wu12,Han12}. 
Indeed, the NDR effect can also be explained with a simple charge
``pinch-off'' argument \cite{Wu12} within a diffusive transport framework,
where the quasi-saturation arises as a particular case. Besides being potentially useful for applications in digital electronics (e.g. memory cells and clock generators), NDR is of particular interest as a means to engineer the current saturation for analog applications.

In this work, we focus on NDR in short-channel graphene transistors operating in the ballistic regime. This phenomenon has been predicted by previous quantum transport 
studies \cite{Dragoman07,NamDo08,Zhao11}, but in such works the origin of NDR was not completely elucidated,
in particular with regard to the effect of contacts and self-consistent electrostatics,
which we found to play an important role in the operating regimes of NDR.
It is the purpose of this work, which is an extension of \cite{Grassi12}, to clarify the origin of NDR in ballistic GFETs and provide guidance to future experiments.

The paper is organized as follows. Section~\ref{sec_review_saturation} reviews the current interpretations of the quasi-saturation in long- and short-channel GFETs, which prepares the ground for understanding the mechanism behind NDR. Then, Section~\ref{sec_models} describes  our simulation models for GFETs. The results are shown in Section~\ref{sec_results}, followed by a discussion in Section~\ref{sec_discussion}. Conclusions are finally drawn in Section~\ref{sec_conclusions}.

\IEEEpubidadjcol

\section{Review of quasi-saturation and NDR in long- vs. short-channel GFETs} \label{sec_review_saturation}

For long-channel GFETs, the phenomenon of quasi-saturation and NDR can be explained using drift-diffusion models, which assume semiclassical diffusive transport \cite{Meric08,Thiele10,Jimenez11}. The drain current $I$ is given by
\begin{equation}
I = q \left[n(x)+p(x)\right] v(x) W ,
\end{equation}
where $q$ is the electronic charge, $n(x)$ and $p(x)$ the sheet concentration of electrons and holes respectively, $v(x)$ the common drift velocity along the transport direction, and $W$ the channel width. 
The output conductance then consists of two contributions:
\begin{equation} \label{eq_gd_diffusive}
g_d = \frac{\partial I}{\partial V_\mathrm{DS}} \propto \frac{\partial \left[n(x)+p(x)\right]}{\partial V_\mathrm{DS}} v(x) + [n(x)+p(x)] \frac{\partial v(x)}{\partial V_\mathrm{DS}} .
\end{equation}
While $v(x)$ generally increases with $V_\mathrm{DS}$ due to the increasing bending of the quasi-Fermi level $E_F(x)$, the sum $n(x)+p(x)$ instead is decreasing when $E_F(x)$ approaches and eventually crosses the Dirac point in the channel $E_d$ at the drain side. This charge ``pinch-off'' is the effect used to explain the quasi-saturation or even NDR \cite{Wu12} in the diffusive regime. 
It is also possible that velocity saturation, due to scattering with substrate polar phonons, contribute to quasi-saturation \cite{Meric08,PA10} and NDR \cite{Han12}, since it implies $\partial v(x)/\partial V_\mathrm{DS} \rightarrow 0$ in the r.h.s. of Eq.~\ref{eq_gd_diffusive}.

For ballistic GFETs instead, quasi-saturation can be understood using the Landauer formalism \cite{Koswatta11}. Assuming for simplicity the zero-temperature approximation, the energy window for transport is the one between the Fermi levels $\mu_D$ and $\mu_S$ within the drain and source regions, respectively. The current is then given by
\begin{equation}
I = \frac{2 q}{h} \int_{\mu_D}^{\mu_S} M(E) \mathrm{d}E ,
\end{equation}
where $h$ is Planck's constant, $M(E) = 2 W \left| E-E_d \right| / (\pi \hbar v_F)$ the number of propagating modes in the graphene channel at energy $E$, $\hbar=h/(2\pi)$, and $v_F$ the graphene Fermi velocity. In this case, the output conductance is
\begin{equation} \label{eq_gd_ballistic}
g_d = \frac{\partial I}{\partial V_\mathrm{DS}} \propto q M(\mu_D) + \int_{\mu_D}^{\mu_S} \frac{\partial M(E)}{\partial V_\mathrm{DS}} \mathrm{d}E .
\end{equation}
$M(E)$ is shifted up or down in energy by varying $E_d$, which in turn is determined by the gate electrostatics. Assuming that $E_d$ does not depend on $V_\mathrm{DS}$, the second term in the r.h.s. of Eq.~\ref{eq_gd_ballistic} is zero. Consequently, it can be seen that the current tends to saturate when $\mu_D$ approaches $E_d$ because $M(E_d) = 0$ (here we neglect the graphene minimum conductivity \cite{TTT06,SLL11}). 
The sign of $g_d$ in Eq.~\ref{eq_gd_ballistic} can be negative only if $\partial M(E)/\partial V_\mathrm{DS}$ is negative,
which is not obvious from this simple picture.
In the following, we will generalize this simple model
to include the effects of contacts and 
self-consistent electrostatics, and show that 
$\partial M(E)/\partial V_\mathrm{DS}$ can indeed be negative
within the energy window for transport.

\section{Models} \label{sec_models}

\begin{figure}[!t]
\centering
\includegraphics[scale=0.3]{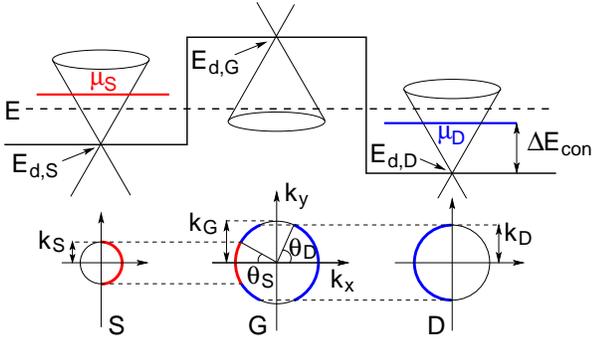}
\caption{Semianalytical model: profile of the Dirac point $E_d(x)$ (top) and population of states in $k$-space (bottom) corresponding to the indicated energy $E$. The cones represent the graphene dispersion relation $E(\vec{k}) = E_d \pm\hbar v_F |\vec{k}|$. States populated with the Fermi function of the source/drain are indicated with a thick red/blue (light-gray/dark-gray) line.}
\label{fig_EdG_with_contacts}
\end{figure}

In this section, we describe in detail the semianalytical model and the numerical model which are used to investigate the device behavior. In the considered GFETs the source and drain regions are assumed to be made of a graphene layer with a metal layer deposited on top.

The semianalytical model assumes a simple ideal square potential barrier, as shown in Fig.~\ref{fig_EdG_with_contacts}-top, where $E_{d,S/D/G}$ is the energy of the Dirac point in the source/drain/channel region. This is a good approximation for the case of self-aligned contacts as shown later by comparison with our numerical treatment. The related metal-induced doping \cite{GKBK08} is introduced through a fixed difference $\Delta E_{\mathrm{con}}$ between the Fermi level and the Dirac point in the source and drain regions (Fig.~\ref{fig_EdG_with_contacts}-top). 

The transport model is composed of the following equations for the electron and hole concentration in the channel region and for the drain current:
\begin{align}
n & = \int_{E_{d,G}}^{\infty} \mathrm{d}E \left[ D_L(E) f_S(E) + D_R(E) f_D(E) \right], \label{eq_n} \\
p & = \int_{-\infty}^{E_{d,G}} \mathrm{d}E \left[ D_L(E) \left(1-f_S(E)\right) + D_R(E) \left(1-f_D(E)\right) \right], \label{eq_p} \\
I & = \frac{2 q}{h} \int_{-\infty}^{\infty} \mathrm{d}E \, M(E) [f_S(E) -f_D(E)] , \label{eq_current}
\end{align}
where $D_{L/R}(E)$ is the density of states (DOS) in the channel at energy $E$ relative to injection from source/drain and $f_{S/D}(E)$ the contact Fermi distribution with Fermi level $\mu_{S/D}$. In turn, the model for $D_L(E)$, $D_R(E)$, and $M(E)$ is given by
\begin{align}
D_L(E) & = \frac{| E - E_{d,G} |}{(\pi \hbar v_F)^2} 2 \left[ \theta_S + (\theta_S - \theta_D ) \vartheta  (\theta_S - \theta_D) \right] , \label{eq_Ds}\\
D_R(E) & = \frac{| E - E_{d,G} |}{(\pi \hbar v_F)^2} 2 \left[ \theta_D + (\theta_D - \theta_S ) \vartheta  (\theta_D - \theta_S) \right] , \label{eq_Dd}\\
M(E) & = \min \left\{ M_S(E), M_D(E), M_G(E)\right\} , \label{eq_M}
\end{align}
where $\vartheta$ is the Heaviside step function and
\begin{equation}
M_{S/D/G}(E) = \frac{2 W}{\pi}k_{S/D/G} \, . \label{eq_M_2}
\end{equation}
The numerical value for $v_F$ is set using the equation $v_F = (3/2) a_\mathrm{CC} |t| / \hbar$, in which $a_\mathrm{CC} = 1.42$~\AA{} is the carbon-carbon distance in graphene and $t=-2.7$~eV the tight-binding parameter describing hopping between nearest neighbor $p_z$ orbitals. The quantities $k_{S/D/G}$ and $\theta_{S/D}$ are defined as (their physical meaning is illustrated in Fig.~\ref{fig_EdG_with_contacts}-bottom)
\begin{align}
k_{S/D/G} & = \frac{\left| E - E_{d,S/D/G} \right|}{\hbar v_F} , \label{eq_kSDG}\\
\theta_{S/D} & = \left\{
\begin{array}{ll}
\sin^{-1} (k_{S/D}/k_G) & \mbox{if $k_{S/D} < k_G$,} \\
\pi/2 & \mbox{otherwise.}
\end{array}
\right. \label{eq_thetaSD}
\end{align}

The Dirac point in the channel $E_{d,G}$ is self-consistently computed with $n$ and $p$ through a plane-capacitor model which accounts for electrostatics:
\begin{equation}
q (n-p) = C_{\mathrm{ox}} \left( - \mu_S/q + V_{\mathrm{GS}} + E_{d,G}/q \right) , \label{eq_electrostatics}
\end{equation}
where $C_{\mathrm{ox}}$ is the gate oxide capacitance and a zero workfunction difference is assumed between gate and graphene.

The model in Eqs.~\ref{eq_Ds}--\ref{eq_thetaSD} corresponds to the solution of the ballistic Boltzmann equation in the channel region assuming energy and transverse momentum conservation at the two junctions and including Klein tunneling \cite{KNG06} with tunneling probability equal to 1. As an example, Fig.~\ref{fig_EdG_with_contacts}-bottom shows the distribution function in $k$-space corresponding to the potential in Fig.~\ref{fig_EdG_with_contacts}-top and at the indicated energy $E$. The red/blue (light-gray/dark-gray) color represents $f_{S/D}(E)$. The plot can be understood by assuming that the transmission probability across each junction, for an incident electron with transverse momentum $k_y$, is either $1$, if states with the same $k_y$ are available on the other side of the junction, or $0$ otherwise. In the figure, electrons from the source (red or light gray) are perfectly transmitted through both junctions, thus populating only rightward propagating states in the channel (note that the group velocity is opposite to $\vec{k}$ for states in the valence band). The ones from the drain (blue or dark gray) enter the channel with probability one; at the source-channel junction, they are either perfectly transmitted if $|k_y| < k_S$ or totally reflected if $|k_y| > k_S$, thus populating both left- and rightward propagating states.

We highlight the fact that the current contribution at a given energy is determined by the region where the Fermi surface has the smallest radius (Eq.~\ref{eq_M}), which means a transport bottleneck effect due to the series of graphene junctions. This model for $M(E)$ was first discussed in \cite{Low09} and was also used to describe NDR in single p-n junction devices \cite{Fiori11}. It is worth noting that, if $k_S, k_D > k_G$ (i.e., if the number of modes in the contacts is larger than in the channel), the same model for charge and current as in \cite{Koswatta11} is recovered.

To benchmark the semianalytical model, we use an atomistic full-quantum code \cite{Imperiale10}, based on the self-consistent solution of the tight-binding (TB) NEGF and 3D Poisson equations and optionally including graphene acoustic phonon (AP) and optical phonon (OP) scattering. The source and drain regions are treated as in the semianalytical model with a fixed $\Delta E_{\mathrm{con}}$, semi-infinite extensions, and zero underlap between the source and drain contacts and the gate (as in \cite{Zhao11}). Both ballistic simulations and simulations with phonon scattering have been performed; in the latter case, we use the parameters  $D_{\omega,\mathrm{AP}} = 0.03$ $\mathrm{eV^2}$, $D_{\omega,\mathrm{OP}} = 0.027$ $\mathrm{eV^2}$, and $\hbar \omega_\mathrm{OP} = 160$ meV, whose definitions can be found in \cite{Yoon11}.

So far, we have assumed that the source and drain regions are described by the same conical electronic dispersion relation as the channel (Eq.~\ref{eq_kSDG}). In reality, the graphene DOS in the contacted regions is broadened due to the coupling with the metal contacts, so that a finite DOS (and thus a finite current injection) is induced at the Dirac point. In the following, we study separately the effect of contact-induced energy broadening. Regarding the NEGF code, we include the broadening as a constant imaginary diagonal self-energy $-\mathrm{i} \Delta$ for the source and drain regions \cite{Zhao11}. We have verified that the resulting DOS in the source/drain region, $D_{S/D}(E)$, can be well reproduced by the formula
\begin{equation}
D_{S/D}(E) = 2 \frac{\sqrt{\left(E-E_{d,S/D}\right)^2 + \widetilde{\Delta}^2}}{\pi \left(\hbar v_F\right)^2}
\end{equation}
where $\widetilde{\Delta}$ is a fitting parameter. Assuming the same relation between $M_{S/D}$ and $D_{S/D}$ as in the case without broadening,
\begin{equation}
M_{S/D}(E) = D_{S/D}(E) \hbar v_F W,
\end{equation}
we get an effective dispersion relation
\begin{equation}
k_{S/D} = \frac{\pi}{2 W} M_{S/D}(E) = \frac{\sqrt{\left(E-E_{d,S/D}\right)^2 + \widetilde{\Delta}^2}}{\hbar v_F} , \label{eq_kSD_Delta}
\end{equation}
which we use in place of Eq.~\ref{eq_kSDG} for $k_{S/D}$ to capture the effect of energy broadening within the semianalytical model.

\section{Results} \label{sec_results}

We consider n-type doped source and drain regions. Unless stated otherwise, we assume the values $\Delta E_\mathrm{con} = 0.4$~eV, $\widetilde{\Delta} = \Delta = 0$, and an equivalent oxide thickness (EOT) of the gate dieletric of 0.5~nm. All the results presented in the following are at room temperature. With $V_\mathrm{GS} = 0$ and at equilibrium, the channel Dirac point is aligned with the source and drain Fermi levels and the channel is intrinsic; by applying a positive (negative) $V_\mathrm{GS}$, the bands in the channel are shifted down (up) thus creating an n-n-n (n-p-n) double junction. We explore in the following both the n-n-n and n-p-n bias conditions.

\subsection{Quasi-saturation in n-n-n structure with $V_\mathrm{DS} > 0$} \label{sec_saturation}

\begin{figure}[!t]
\centering
\includegraphics[scale=0.3]{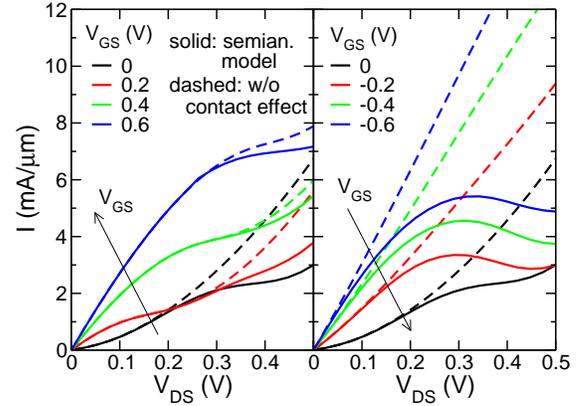}
\caption{Normalized output characteristics for $V_{\mathrm{GS}} \ge 0$, i.e. n-n-n configuration (left), and for $V_{\mathrm{GS}} \le 0$, i.e. n-p-n configuration (right), from the semianalytical model (solid line) and the one obtained by assuming $k_S, k_D > k_G$ (dashed line).}
\label{fig_output}
\end{figure}

In Fig.~\ref{fig_output}-left, we plot the output characteristics for $V_{\mathrm{GS}} \ge 0$ (n-type channel) from the semianalytical model and the one obtained by setting $k_S, k_D > k_G$ and thus $\theta_S = \theta_D = \pi/2$ in Eqs.~\ref{eq_Ds}--\ref{eq_M_2} (cfr.~\cite{Koswatta11}). It can be seen that the two models give similar results at large $V_{\mathrm{GS}}$, both predicting the quasi-saturation behavior discussed above. This means that the mode reflection at the two junctions has no significant effect in this bias condition. However, at lower $V_{\mathrm{GS}}$, the two models depart significantly for large $V_\mathrm{DS}$. Indeed, at small $V_{\mathrm{GS}}$ and large $V_{\mathrm{DS}}$, the channel doping is actually turned p-type by the drain contact and the transport regime is similar to the n-p-n case discussed below.

\subsection{NDR in n-p-n structure with $V_\mathrm{DS} > 0$} \label{sec_ndr_positiveVd}

The output characteristics for $V_{\mathrm{GS}} \le 0$ (p-type channel) are shown in Fig.~\ref{fig_output}-right. While the model neglecting the finite number of modes in the source and drain predicts a monotonically increasing current, the one proposed here clearly gives NDR.

\begin{figure}[!t]
\centering
\includegraphics[scale=0.3]{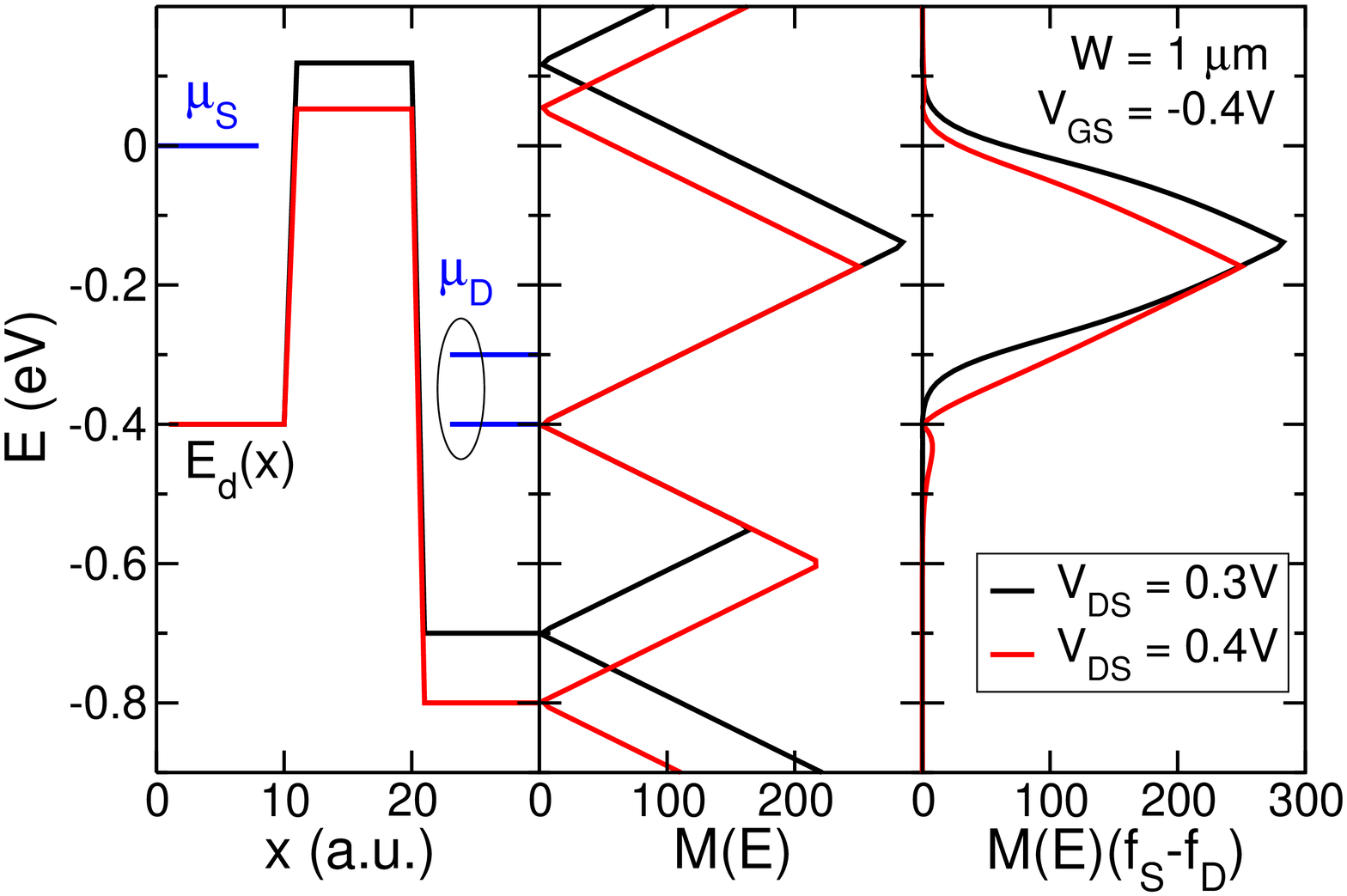}
\caption{Dirac point profile (left), number of propagating modes versus energy (center), and current spectrum (right) from the semianalytical model, for two different $V_{\mathrm{DS}}$ at $V_{\mathrm{GS}}=-0.4$~V.}
\label{fig_spectra}
\end{figure}

The origin of the NDR effect is explained by looking at Fig.~\ref{fig_spectra}, which compares the band profile, number of modes, and current spectrum (integrand in Eq.~\ref{eq_current}) obtained at two different $V_\mathrm{DS}$ biases along the $V_{\mathrm{GS}} = -0.4$~V curve. NDR is the combination of two effects. First, for energies close to $E_{d,S}$, $M(E)$ is limited by the number of modes in the source: in particular, for $E=E_{d,S}$ we have $M(E) = 0$, and this leads to the quasi-saturation behavior of the current for $\mu_D$ approaching $E_{d,S}$, as already pointed out in \cite{Zhao11} and mathematically represented (in the zero-temperature approximation) by the first term in the r.h.s. of Eq.~\ref{eq_gd_ballistic}. Secondly, by lowering $\mu_D$, the flux of electrons injected from the drain into the channel is reduced, while the flux from the source is kept fixed, causing a hole pile-up in the channel: due to the electrostatic feedback, $E_{d,G}$ is lowered. Since for energies close to $E_{d,G}$, $M(E)$ is limited by the number of modes in the channel, the lowering of $E_{d,G}$ causes a decrease in $M(E)$ for a portion of the energy range within $\mu_S$ and $\mu_D$, leading to a decrease of the current rather than a saturation, as expressed (in the zero-temperature approximation) by the second term in the r.h.s of Eq.~\ref{eq_gd_ballistic}. The fact that the decrease in $M(E)$ is not fully compensated by the larger $(\mu_S-\mu_D)$ is confirmed by the plot in Fig.~\ref{fig_spectra}-right, where the area under the red curve (current at higher $V_\mathrm{DS}$) is slightly smaller than the area under the black curve (current at lower $V_\mathrm{DS}$).

We highlight the necessity of two Dirac points, one in the channel and the other one in either the source or the drain, for NDR to be possible instead of saturation. Also, we note that a similar explanation for NDR was given in \cite{NamDo08}, even if a simple, not self-consistent model for barrier lowering was used (shift of $E_{d,G}$ by $-V_\mathrm{DS}/2$ with respect to the value at $V_\mathrm{DS}=0$).

\begin{figure}[!t]
\centering
\includegraphics[scale=0.3]{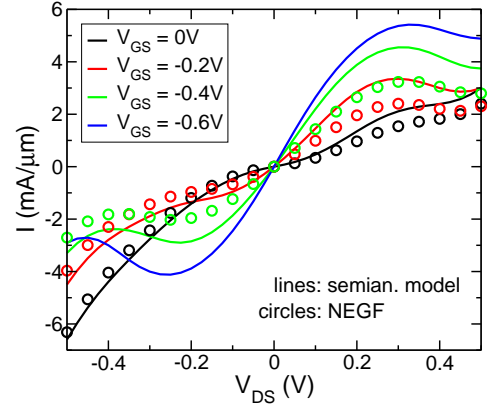}
\caption{Normalized output characteristics for $V_{\mathrm{GS}} \le 0$ and for both positive and negative $V_{\mathrm{DS}}$ from the semianalytical model and ballistic NEGF simulations.}
\label{fig_output_neg_full}
\end{figure}

\begin{figure}[!t]
\centering
\includegraphics[scale=0.3]{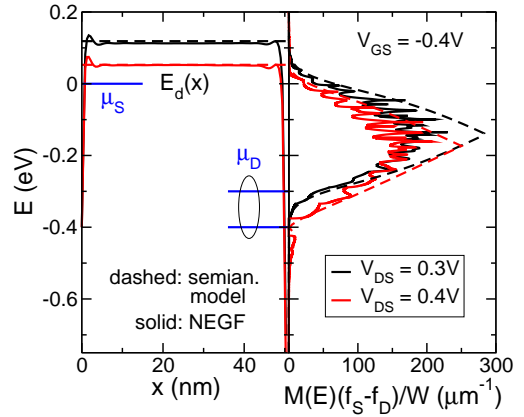}
\caption{Dirac point profile (left) and current spectrum (right) from the semianalytical model and from ballistic NEGF simulations for the same two biases as in Fig.~\ref{fig_spectra}.}
\label{fig_spectra_comparison}
\end{figure}

The qualitative shape of the output characteristics is confirmed by ballistic NEGF simulations considering a 50-nm-channel length (Fig.~\ref{fig_output_neg_full}: see positive $V_\mathrm{DS}$ axis, the results for negative $V_\mathrm{DS}$ being discussed in the following section).
In Fig.~\ref{fig_spectra_comparison} we plot the band profiles and current spectra obtained from the two models. In the ballistic NEGF formalism, the number of propagating modes $M(E)$ in the current expression is replaced by the transmission function $T(E) = \sum_{j=1} T_j(E)$, where $T_j(E)$ is the transmission probability from source to drain of mode $j$ \cite{Datta95}. From the figure, it can be seen that the assumption of square potential barrier is well justified and that the barrier lowering is similar; the lower current spectrum in the NEGF case can be explained with wavefunction mismatch at the junctions, causing $T_j(E)<1$ for propagating modes even in the case of an abrupt potential step \cite{KNG06,Low09}.

Finally, we have investigated the effect of scattering due to graphene longitudinal acoustic and optical phonons and found that, at this channel length, it is too weak to affect the current (see Fig.~6 in \cite{Grassi12}).

\subsection{NDR in n-p-n structure with $V_\mathrm{DS} < 0$} \label{sec_ndr_negativeVd}

\begin{figure}[!t]
\centering
\includegraphics[scale=0.3]{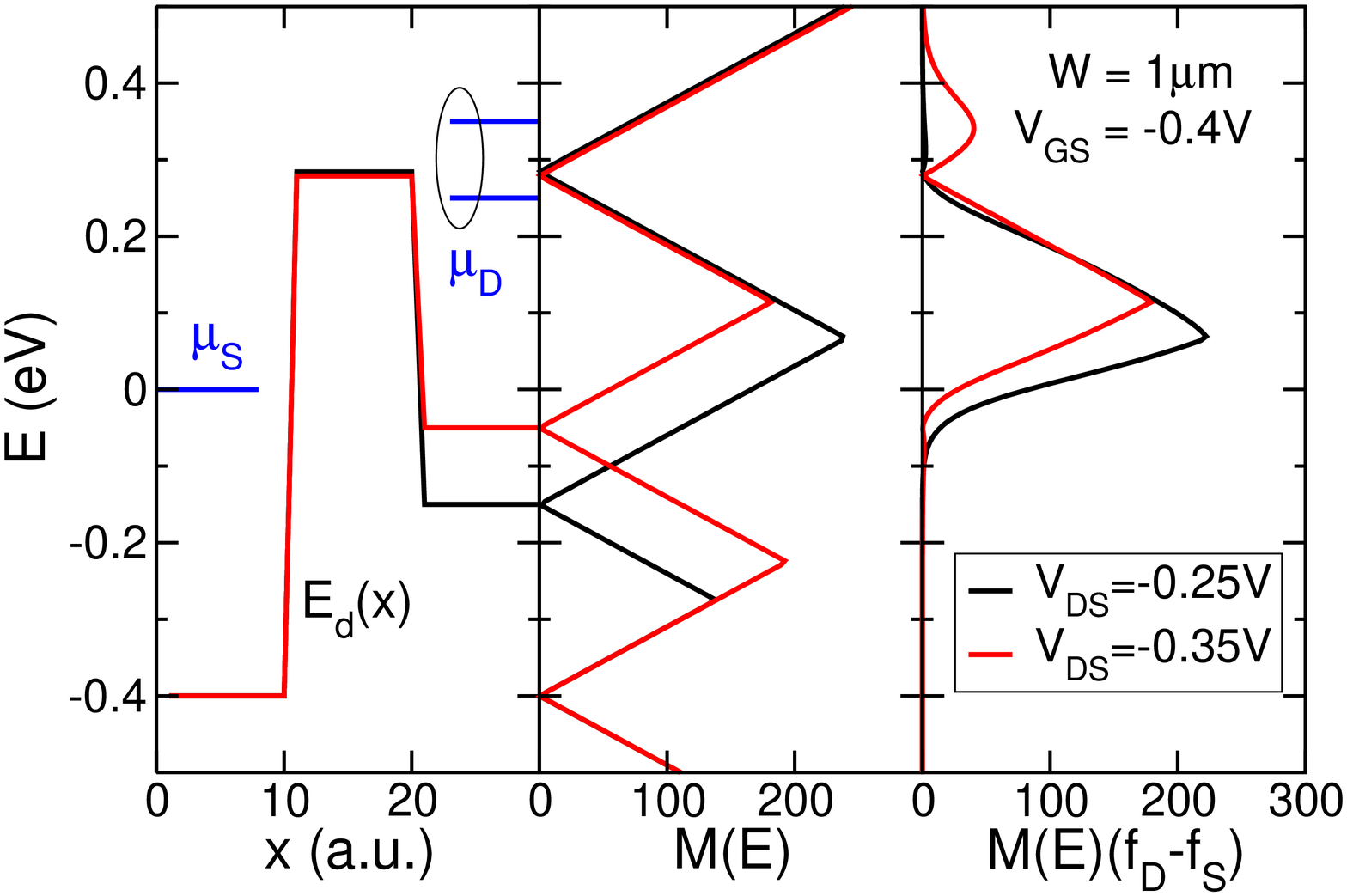}
\caption{Dirac point profile (left), number of propagating modes versus energy (center), and current spectrum (right) from the semianalytical model, for two different negative $V_{\mathrm{DS}}$ at $V_{\mathrm{GS}}=-0.4$~V.}
\label{fig_spectra_negativeVd}
\end{figure}

We have also studied the device behavior when the polarity of the drain voltage is reversed (Fig.~\ref{fig_output_neg_full}). Another NDR effect, not previously reported, is observed at $V_\mathrm{DS}<0$ in both the results of the semianalytical model and NEGF simulations.

To help understand the origin of the phenomenon, we plot in Fig.~\ref{fig_spectra_negativeVd} the band profile and spectra for two negative $V_\mathrm{DS}$ values along the $V_\mathrm{GS}=-0.4$~V curve. The mode bottleneck induced by the Dirac point at $E=E_{d,G}$ is responsible for the current saturation as $\mu_D$ approaches $E_{d,G}$, since it gives $M(\mu_D) \rightarrow 0$ in Eq.~\ref{eq_gd_ballistic}. At the same time, when $\mu_D$ is raised, the mode bottleneck at $E=E_{d,D}$ causes a decrease of $M(E)$ within the energy window between $\mu_S$ and $\mu_D$, due to the rigid shift of $E_{d,D}$ with $\mu_D$. This corresponds to $\partial M(E) / \partial V_\mathrm{DS} < 0$ in Eq.~\ref{eq_gd_ballistic}, so that NDR rather than saturation occurs. The reason why $E_{d,G}$ is almost unchanged when varying $V_\mathrm{DS}$, as opposed to the $V_\mathrm{DS}>0$ case, is related to the vanishing DOS in the channel for $E=E_{d,G}$ (cfr. Eq.~\ref{eq_Dd}), so that the charge variation in the channel induced by the lifting of $\mu_D$ tends to zero for $\mu_D$ approaching $E_{d,G}$.
 
We conclude that the mechanisms for NDR at the two drain voltage polarities are very similar and essentially related to the combined effect of two Dirac points, one in the channel and the other in either the source or the drain, in limiting the current. However, NDR at $V_\mathrm{DS}<0$ does not involve an electrostatic feedback in the channel as opposed to its counterpart at $V_\mathrm{DS}>0$. We note that, in the former case, since $E_{d,G}$ does not move with respect to $\mu_S$ at fixed $V_\mathrm{GS}$, the NDR effect reported here is essentially the same as the one predicted for single p-n junctions in \cite{Fiori11}.

\begin{figure}[!t]
\centering
\includegraphics[scale=0.3]{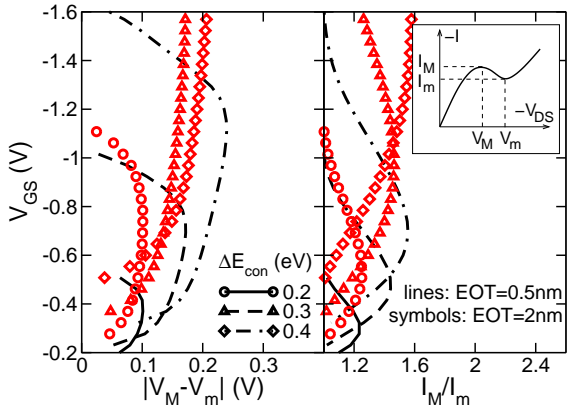}
\caption{Dependence on $V_\mathrm{GS}$, $\Delta E_\mathrm{con}$, and EOT of the voltage swing (left) and peak-to-valley current ratio (right) of NDR at $V_\mathrm{DS} < 0$. The results are from the semianalytical model. The inset shows the definition of $V_M$, $V_m$, $I_M$, and $I_m$.}
\label{fig_doping}
\end{figure}

While the optimization of NDR is beyond the scope of this work, we just show in Fig.~\ref{fig_doping} how NDR at $V_\mathrm{DS} < 0$ can be modulated by varying the electrostatic doping of the contacts, i.e. $\Delta E_\mathrm{con}$, and EOT. From the figure, it can be seen that both the voltage swing $|V_M-V_m|$ and the peak-to-valley current ratio $I_M/I_m$ (the symbols are defined in the inset of Fig.~\ref{fig_doping}) can be somewhat enhanced by increasing the contact doping, which also enlarges the $V_\mathrm{GS}$ window where the phenomenon occurs. The use of a thicker gate dielectric just shifts the onset of NDR to higher $V_\mathrm{GS}$ values and, at the same time, enlarges the $V_\mathrm{GS}$ window.

\subsection{Effect of energy broadening in the source and drain regions}

\begin{figure}[!t]
\centering
\includegraphics[scale=0.3]{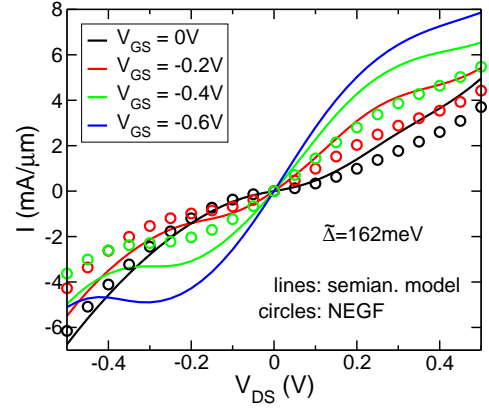}
\caption{Normalized output characteristics for $V_{\mathrm{GS}} \le 0$ from the semianalytical model and ballistic NEGF simulations including energy broadening in the source and drain.}
\label{fig_output_neg_full_Delta}
\end{figure}

To study the effect of energy broadening due to metal-graphene coupling in the source and drain regions, we set $\Delta=50$~meV in the NEGF code and $\widetilde{\Delta} = 162$~meV in the semianalytical model. The former value is taken from \cite{Zhao11}; the latter has been fitted to provide the same value of DOS at the Dirac point, $D_{S/D}(E_{d,S/D}) \approx 3 \times 10^{13} \textrm{cm}^{-2} \textrm{eV}^{-1}$, as in the NEGF case. The resulting output characteristics are plotted in Fig.~\ref{fig_output_neg_full_Delta}. The two methods give qualitatively similar results. At $V_\mathrm{DS}>0$, NDR disappears as already observed in \cite{Zhao11}. However, we find here that quasi-saturation is still possible at $V_\mathrm{DS}<0$, according to the NEGF model. The experimental verification of the effect should thus be easier in the $V_\mathrm{DS}<0$ case.

\subsection{Effect of contact resistance}

\begin{figure}[!t]
\centering
\includegraphics[scale=0.3]{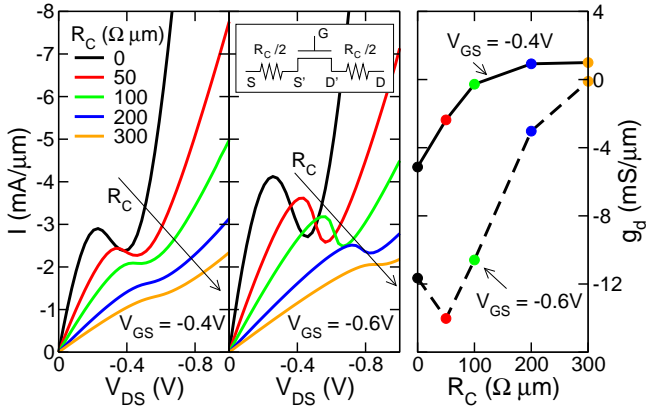}
\caption{Left/center: normalized output characteristics from the semianalytical model for different values of the contact resistance $R_C$, at $V_\mathrm{GS}=-0.4$~V (left) and $V_\mathrm{GS}=-0.6$~V (center). $R_C$ is split equally between the source and drain terminals as shown in the inset. Right: corresponding output conductance at the inflection point, as a function of $R_C$.}
\label{fig_Rc}
\end{figure}

We finally consider the effect of a series contact resistance $R_C$ equally split between the source and drain terminals (inset of Fig.~\ref{fig_Rc}). The semianalytical model is modified accordingly, by replacing the Fermi level $\mu_{S/D}$ entering the equations for $f_{S/D}$ and $E_{d,S/D}$ with the Fermi level $\mu_{S'/D'}$ of the intrinsic source/drain. The latter is calculated self-consistently with the current $I$ through the resistor equation
\begin{equation}
\mu_{S'/D'} = \mu_{S/D} \mp q \frac{R_C}{2} I ,
\end{equation}
where the upper/lower sign holds for source/drain. The $I$ vs. $V_\mathrm{DS}$ characteristics for different values of $R_C$ and two values of $V_\mathrm{GS}$ are shown in Fig.~\ref{fig_Rc}, together with the extracted values of $g_d$ at the inflection point of each curve (defined as the $V_\mathrm{DS}$ voltage where $\partial^2 I / \partial V_\mathrm{DS}^2$ changes its sign). It can be seen that NDR tends to disappear with increasing $R_C$ due to the increasing voltage drop across the two resistors. Already at $R_C=300$~$\Omega\cdot\mu$m, which is a typical experimental value \cite{Wu12}, the output curve resembles the linear characteristic of a resistor, for both values of the gate voltage considered, indicating that contact resistance is a major problem for the operation of short-channel devices which must be addressed in future developments. Progress has recently been made in reducing $R_C$ below $100$~$\Omega\cdot\mu$ \cite{Moon12}.

\section{Discussion} \label{sec_discussion}

The analysis above indicates that, given a specific type of doping of the contacted graphene regions, the gate voltage needs to be biased such that an n-p-n or p-n-p double junction is formed, in order for NDR in ballistic GFETs to be possible. In addition, NDR is expected to be more likely for drain voltages with the same polarity as $V_\mathrm{GS}$, due to the higher robustness of NDR against energy broadening in this case. It is interesting to compare these findings with the actual bias conditions employed in experiments to observe NDR.

We note that the long-channel devices in \cite{Wu12,Han12} were biased in an n-n-n or p-p-p configuration thus ruling out our interpretation of the phenomenon. Indeed, for channel lengths of the order of $1 \mu$m, scattering is expected to cause significant momentum and energy relaxation and charge inhomogeneity along the channel, so that the drift-diffusion interpretation discussed in Section~\ref{sec_review_saturation} seems more appropriate.

On the other hand, in the Supporting Information of \cite{Wu12}, results for GFETs with channel lengths of about $100$~nm have been reported too. Some of these devices show NDR for gate voltages around $0$~V: this biasing could be the counterpart case of the one mentioned at the end of Section~\ref{sec_saturation}, (incompletely formed p-type channel, p-type contacts, and $V_\mathrm{DS} < 0$), so that our explanation of NDR could apply in this case. However, further experimental evidence is needed to prove this hypothesis (for example by exploring all the $V_\mathrm{GS}$ and $V_\mathrm{DS}$ polarities).

\section{Conclusions} \label{sec_conclusions}
Through a semianalytical model and detailed quantum simulations, we have clarified the nature and bias conditions for NDR in short-channel GFETs. The origin of the phenomenon is attributed to the transport-mode bottleneck induced by the graphene Dirac point. The combined effect of two Dirac points, one in the channel and the other in either the source or the drain, is necessary for NDR to occur instead of quasi-saturation. This is verified in the n-p-n or p-n-p configuration, for both polarities of $V_\mathrm{DS}$. It is found that a large doping concentration of the contacts enhances NDR, although the maximum achievable peak-to-valley ratio is limited to about 1.6. In the presence of energy broadening due to the metal-graphene coupling in the source and drain regions, NDR disappears at one $V_\mathrm{DS}$ polarity, but quasi-saturation is still attainable at the other one. It is also found that contact resistance at typical experimental values suppresses NDR, representing a major obstacle for the verification of the phenomenon in experiments. 

The NDR mechanism could offer new possibilities for the optimization of the saturation behavior of the output characteristics of analog GFETs. The semianalytical model presented here, providing a good physical insight of NDR, could be a useful simulation tool for such an optimization study.

\appendices


\section*{Acknowledgment}

The authors would like to thank Dr. Y. Wu, IBM T. J. Watson Research Center, Yorktown Heights, NY, for fruitful discussions about the interpretation of the results.

\ifCLASSOPTIONcaptionsoff
  \newpage
\fi



\bibliographystyle{IEEEtran}
\bibliography{IEEEabrv,mybibfile}

\begin{thebibliography}{10}
\providecommand{\url}[1]{#1}
\csname url@samestyle\endcsname
\providecommand{\newblock}{\relax}
\providecommand{\bibinfo}[2]{#2}
\providecommand{\BIBentrySTDinterwordspacing}{\spaceskip=0pt\relax}
\providecommand{\BIBentryALTinterwordstretchfactor}{4}
\providecommand{\BIBentryALTinterwordspacing}{\spaceskip=\fontdimen2\font plus
\BIBentryALTinterwordstretchfactor\fontdimen3\font minus
  \fontdimen4\font\relax}
\providecommand{\BIBforeignlanguage}[2]{{%
\expandafter\ifx\csname l@#1\endcsname\relax
\typeout{** WARNING: IEEEtran.bst: No hyphenation pattern has been}%
\typeout{** loaded for the language `#1'. Using the pattern for}%
\typeout{** the default language instead.}%
\else
\language=\csname l@#1\endcsname
\fi
#2}}
\providecommand{\BIBdecl}{\relax}
\BIBdecl

\bibitem{MericIEDM08}
I.~Meric, N.~Baklitskaya, P.~Kim, and K.~L. Shepard, ``{RF} performance of
  top-gated, zero-bandgap graphene field-effect transistors,'' in \emph{Int.
  Electron Devices Meeting Tech. Dig.}, 2008, pp. 1--4.

\bibitem{Liao10}
L.~Liao, Y.-C. Lin, M.~Bao, R.~Cheng, J.~Bai, Y.~Liu, Y.~Qu, K.~L. Wang,
  Y.~Huang, and X.~Duan, ``High-speed graphene transistors with a self-aligned
  nanowire gate,'' \emph{Nature}, vol. 467, no. 7313, pp. 305--308, 2010.

\bibitem{Lin10}
Y.-M. Lin, C.~Dimitrakopoulos, K.~A. Jenkins, D.~B. Farmer, H.-Y. Chiu,
  A.~Grill, and P.~Avouris, ``100-{GHz} transistors from wafer-scale epitaxial
  graphene,'' \emph{Science}, vol. 327, no. 5966, p. 662, 2010.

\bibitem{Wu11}
Y.~Wu, Y.~ming Lin, A.~A. Bol, K.~A. Jenkins, F.~Xia, D.~B. Farmer, Y.~Zhu, and
  P.~Avouris, ``High-frequency, scaled graphene transistors on diamond-like
  carbon,'' \emph{Nature}, vol. 472, no. 7341, pp. 74--78, 2011.

\bibitem{Schwierz10}
F.~Schwierz, ``Graphene transistors,'' \emph{Nature Nanotechnology}, vol.~5,
  pp. 487--496, 2010.

\bibitem{Lin11}
Y.-M. Lin, A.~Valdes-Garcia, S.-J. Han, D.~B. Farmer, I.~Meric, Y.~Sun, Y.~Wu,
  C.~Dimitrakopoulos, A.~Grill, P.~Avouris, and K.~A. Jenkins, ``Wafer-scale
  graphene integrated circuit,'' \emph{Science}, vol. 332, no. 6035, pp.
  1294--1297, 2011.

\bibitem{Meric08}
I.~Meric, M.~Y. Han, A.~F. Young, B.~Ozyilmaz, P.~Kim, and K.~L. Shepard,
  ``Current saturation in zero-bandgap, top-gated graphene field-effect
  transistors,'' \emph{Nature Nanotechnology}, vol.~3, pp. 654--659, 2008.

\bibitem{Bai11}
J.~Bai, L.~Liao, H.~Zhou, R.~Cheng, L.~Liu, Y.~Huang, and X.~Duan, ``Top-gated
  chemical vapor deposition grown graphene transistors with current
  saturation,'' \emph{Nano Letters}, vol.~11, no.~6, pp. 2555--2559, 2011.

\bibitem{Han11}
S.-J. Han, K.~A. Jenkins, A.~V. Garcia, A.~D. Franklin, A.~A. Bol, and
  W.~Haensch, ``High-frequency graphene voltage amplifier,'' \emph{Nano
  Letters}, vol.~11, no.~9, pp. 3690--3693, 2011.

\bibitem{Thiele10}
S.~A. Thiele, J.~A. Schaefer, and F.~Schwierz, ``Modeling of graphene
  metal-oxide-semiconductor field-effect transistors with gapless large-area
  graphene channels,'' \emph{J. Appl. Phys.}, vol. 107, no.~9, p. 094505, 2010.

\bibitem{Jimenez11}
D.~Jim\'enez and O.~Moldovan, ``Explicit drain-current model of graphene
  field-effect transistors targeting analog and radio-frequency applications,''
  \emph{{IEEE} Trans. Electron Devices}, vol.~58, no.~11, pp. 4049--4052, 2011.

\bibitem{Koswatta11}
S.~O. Koswatta, A.~Valdes-Garcia, M.~B. Steiner, Y.-M. Lin, and P.~Avouris,
  ``Ultimate {RF} performance potential of carbon electronics,'' \emph{{IEEE}
  Trans. Microw. Theory Tech.}, vol.~59, no.~10, pp. 2739--2750, 2011.

\bibitem{Ganapathi12}
K.~Ganapathi, M.~Lundstrom, and S.~Salahuddin, ``Can quasi-saturation in the
  output charactheristics of short-channel graphene field-effect transistors be
  engineered?'' in \emph{Proc. Device Research Conf.}, University Park, PA,
  Jun. 2012, pp. 85--86.

\bibitem{Wu12}
Y.~Wu, D.~B. Farmer, W.~Zhu, S.-J. Han, C.~D. Dimitrakopoulos, A.~A. Bol,
  P.~Avouris, and Y.-M. Lin, ``Three-terminal graphene negative differential
  resistance devices,'' \emph{ACS Nano}, vol.~6, no.~3, pp. 2610--2616, 2012.

\bibitem{Han12}
S.-J. Han, D.~Reddy, G.~D. Carpenter, A.~D. Franklin, and K.~A. Jenkins,
  ``Current saturation in submicrometer graphene transistors with thin gate
  dielectric: Experiment, simulation, and theory,'' \emph{ACS Nano}, vol.~6,
  no.~6, pp. 5220--5226, 2012.

\bibitem{Dragoman07}
D.~Dragoman and M.~Dragoman, ``Negative differential resistance of electrons in
  graphene barrier,'' \emph{Appl. Phys. Lett.}, vol.~90, no.~14, p. 143111,
  2007.

\bibitem{NamDo08}
V.~N. Do, V.~H. Nguyen, P.~Dollfus, and A.~Bournel, ``Electronic transport and
  spin-polarization effects of relativisticlike particles in mesoscopic
  graphene structures,'' \emph{J. Appl. Phys.}, vol. 104, no.~6, p. 063708,
  2008.

\bibitem{Zhao11}
P.~Zhao, Q.~Zhang, D.~Jena, and S.~O. Koswatta, ``Influence of metal-graphene
  contact on the operation and scalability of graphene field-effect
  transistors,'' \emph{{IEEE} Trans. Electron Devices}, vol.~58, no.~9, pp.
  3170--3178, 2011.

\bibitem{Grassi12}
R.~Grassi, T.~Low, A.~Gnudi, and G.~Baccarani, ``Negative differential
  resistance in short-channel graphene {FETs}: semianalytical model and
  simulations,'' in \emph{Proc. Device Research Conf.}, University Park, PA,
  Jun. 2012, pp. 107--108.

\bibitem{PA10}
V.~Perebeinos and P.~Avouris, ``Inelastic scattering and current saturation in
  graphene,'' \emph{Phys. Rev. B}, vol.~81, p. 195442, 2010.

\bibitem{TTT06}
J.~Tworzydlo, B.~Trauzettel, M.~Titov, A.~Rycerz, and C.~W.~J. Beenakker,
  ``Sub-poissonian shot noise in graphene,'' \emph{Phys. Rev. Lett.}, vol.~96,
  p. 246802, 2006.

\bibitem{SLL11}
Y.~Sui, T.~Low, M.~Lundstrom, and J.~Appenzeller, ``Signatures of disorder in
  the minimum conductivity of graphene,'' \emph{Nano Lett.}, vol.~11, p. 1319,
  2011.

\bibitem{GKBK08}
G.~Giovannetti, P.~A. Khomyakov, G.~Brocks, V.~M. Karpan, J.~van~den Brink, and
  P.~J. Kelly, ``Doping graphene with metal contacts,'' \emph{Phys. Rev.
  Lett.}, vol. 101, p. 026803, 2008.

\bibitem{KNG06}
M.~I. Katsnelson, K.~S. Novoselov, and A.~K. Geim, ``Chiral tunnelling and the
  {Klein} paradox in graphene,'' \emph{Nature Phys.}, vol.~2, pp. 620--625,
  2006.

\bibitem{Low09}
T.~Low, S.~Hong, J.~Appenzeller, S.~Datta, and M.~S. Lundstrom, ``Conductance
  asymmetry of graphene p-n junction,'' \emph{{IEEE} Trans. Electron Devices},
  vol.~56, no.~6, pp. 1292--1299, 2009.

\bibitem{Fiori11}
G.~Fiori, ``Negative differential resistance in mono and bilayer graphene p-n
  junctions,'' \emph{{IEEE} Electron Device Lett.}, vol.~32, no.~10, pp.
  1334--1336, 2011.

\bibitem{Imperiale10}
I.~Imperiale, R.~Grassi, A.~Gnudi, S.~Reggiani, E.~Gnani, and G.~Baccarani,
  ``Full-quantum calculations of low-field channel mobility in graphene
  nanoribbon {FETs} including acoustic phonon scattering and edge roughness
  effects,'' in \emph{Proc. ULIS}, Glasgow, UK, Mar. 2010, pp. 57--60.

\bibitem{Yoon11}
Y.~Yoon, D.~E. Nikonov, and S.~Salahuddin, ``Role of phonon scattering in
  graphene nanoribbon transistors: Nonequilibrium {Green's} function method
  with real space approach,'' \emph{Appl. Phys. Lett.}, vol.~98, no.~20, p.
  203503, 2011.

\bibitem{Datta95}
S.~Datta, \emph{Electronic Transport in Mesoscopic Systems}.\hskip 1em plus
  0.5em minus 0.4em\relax Cambridge, UK: Cambridge University Press, 1997.

\bibitem{Moon12}
J.~S. Moon, M.~Antcliffe, H.~C. Seo, D.~Curtis, S.~Lin, A.~Schmitz,
  I.~Milosavljevic, A.~A. Kiselev, R.~S. Ross, D.~K. Gaskill, P.~M. Campbell,
  R.~C. Fitch, K.-M. Lee, and P.~Asbeck, ``Ultra-low resistance ohmic contacts
  in graphene field effect transistors,'' \emph{Appl. Phys. Lett.}, vol. 100,
  no.~20, p. 203512, 2012.

\end{thebibliography}
\end{document}